\documentclass[prl,twocolumn,aps,superscriptaddress]{revtex4}
\usepackage{graphicx}
\usepackage{amsfonts}

\usepackage{amsfonts}
\usepackage{amssymb}
\usepackage{amsmath}
\usepackage{dsfont}
\usepackage{hyperref}
\usepackage{slashed}
\usepackage{empheq}
\usepackage{multirow}
\usepackage{fancyhdr}
\usepackage{appendix}
\usepackage{subfigure}
\usepackage{times}

\DeclareMathOperator{\hz}{h\relax{\kern-.15em}z}
\DeclareMathOperator{\pz}{\psi\relax{\kern-.15em}z}

\usepackage[english]{babel}
\usepackage{amssymb}
\usepackage{amsfonts}
\usepackage{amsmath}
\usepackage{graphicx}
\usepackage{epsfig}
\usepackage{psfrag}
\newcommand{\be}{\begin{equation}} \newcommand{\ee}{\end{equation}}
\newcommand{\bea}{\begin{eqnarray}} \newcommand{\eea}{\end{eqnarray}}
\newcommand{\beann}{\begin{eqnarray*}}  \newcommand{\eeann}{\end{eqnarray*}}
\newcommand{\bfig}{\begin{figure}} \newcommand{\efig}{\end{figure}}
\newcommand{\ba}{\begin{array}} \newcommand{\ea}{\end{array}}
\newcommand{\bcen}{\begin{center}} \newcommand{\ecen}{\end{center}}
\newcommand{\btab}{\begin{tabular}} \newcommand{\etab}{\end{tabular}}

\newcommand{\matt}{\left ( \begin{array}{ccc}}
    \newcommand{\ematt}{\end{array} \right )} \newcommand{\matf}{\left ( \begin{array}{cccc}}
    \newcommand{\ematf}{\end{array} \right )} \newcommand{\vect}{\left ( \begin{array}{c}}
    \newcommand{\evect}{\end{array} \right )}    \def\beqn{\begin{eqnarray}}
 \def\eeqn{\end{eqnarray}}  
     
\newtheorem{Proposition}{Proposition}[section]

\newtheorem{Theorem}{Theorem}[section]
\newtheorem{Lemma}{Lemma}[section]
\newtheorem{Corrolary}{Corrolary}[section]

\newcommand{\bp}{\begin{Proposition}}	\newcommand{\ep}{\end{Proposition}}
\newcommand{\bt}{\begin{Theorem}}	\newcommand{\et}{\end{Theorem}}
\newcommand{\bl}{\begin{Lemma}}		\newcommand{\el}{\end{Lemma}}
\newcommand{\bc}{\begin{Corrolary}}	\newcommand{\ec}{\end{Corrolary}}

\begin{document}


\title{ A Dirty Holographic Superconductor}

\author{D. Are\'an}\email{darean@ictp.it}
\affiliation{International Centre for Theoretical Physics (ICTP), Strada Costiera 11, I 34014 Trieste, Italy}
\affiliation{INFN, Sezione di Trieste, Strada Costiera 11, I 34014 Trieste, Italy}

\author{A. Farahi}\email{aryaf@umich.edu}
\affiliation{Michigan Center for Theoretical Physics, University of Michigan, Ann Arbor, MI 48109, USA}

\author{L. A. Pando Zayas}\email{lpandoz@umich.edu}
\affiliation{Michigan Center for Theoretical Physics, University of Michigan, Ann Arbor, MI 48109, USA}

\author{I. Salazar Landea}\email{peznacho@gmail.com}
\affiliation{ Instituto de F\'\i sica La Plata (IFLP) and Departamento de F\'\i sica Universidad Nacional de La Plata, CC 67,
1900 La Plata, Argentina}
\affiliation{International Centre for Theoretical Physics (ICTP), Strada Costiera 11, I 34014 Trieste, Italy}

\author{A. Scardicchio}\email{ascardic@ictp.it}
\affiliation{International Centre for Theoretical Physics (ICTP), Strada Costiera 11, I 34014 Trieste, Italy}
\affiliation{INFN, Sezione di Trieste, Strada Costiera 11, I 34014 Trieste, Italy}

\date{\today}
\begin{abstract}
We study the effects of disorder on a holographic superconductor by introducing a random chemical potential on the
boundary.  We consider various realizations of disorder and find that the critical temperature for superconductivity
is enhanced.
We also present evidence for a
precise form of renormalization in this system. Namely, when the random chemical potential is characterized by
a  Fourier spectrum of the form $k^{-2\alpha}$ we find that the spectra of the condensate and the charge density
are again power-laws, whose exponents are accurately and universally governed by linear functions of $\alpha$.
\end{abstract}

\pacs{ }

\maketitle


{\it  Introduction.}-- Disorder is a fundamental paradigm in condensed matter physics as it provides a crucial
step away from clean systems towards realistic ones. One of the most striking and ubiquitous manifestations of
disorder in non-interacting quantum systems is the phenomenon of Anderson localization \cite{Anderson58}, where
the conductivity can be completely suppressed by quantum effects. Due to the technical difficulties involved, the study of the interplay between disorder and interactions in quantum systems has seen little progress on the theoretical side. Recently however, in the context of disordered conductors, Basko, Aleiner and  Altshuler took a formidable step forward by presenting compelling evidence in favor of a many-body localized phase, based on an analysis of the perturbation theory in electron-electron interaction to all orders \cite{BAA}. Subsequent works (see \cite{oganesyan2007,pal2010many,Iyer,Buccheri,deluca} and references therein) have confirmed and sharpened the picture of the existence of a phase transition separating the weak and strong interacting limit of electrons in disordered potentials.

The AdS/CFT correspondence provides a natural framework where some strongly coupled field theory systems can be
described in terms of weakly coupled gravity. It is only natural to try to understand the interplay of disorder
and strong interactions in this context. Indeed, there has been a number of discussions along these lines:
\cite{Hartnoll:2007ih, Hartnoll:2008hs,Fujita:2008rs,Ryu:2011vq,Adams:2011rj,Adams:2012yi,Saremi:2012ji}.
In this paper, however, we follow a direct approach of coupling a given operator to a randomly distributed space-dependent source. We essentially translate a typical condensed matter protocol into the AdS/CFT framework.

One particularly important application of disorder is in the context of dirty superconductors which have a rich history in condensed matter physics dating back to the pioneering work of Anderson in 1959 \cite{Anderson59}. For many years Anderson's theorem, stating that superconductivity is insensitive to perturbations that do not destroy time-reversal invariance (pair breaking), provided the central intuition. Critiques to Anderson's argument were raised, for example, in \cite{Kapitulnik,Ma,Maekawa} where the effects of strong localization were considered. More generally, the question of the role of interactions, in particular, the Coulomb interaction in dirty superconductors connote be considered  settled. In view of this situation, it makes sense to consider alternative models where the problem can be analyzed in full detail.

The AdS/CFT correspondence has succeeded in constructing a holographic version of superconductors
\cite{Hartnoll:2008vx,Hartnoll:2008kx}, for comprehensive reviews see \cite{Hartnoll:2009sz,Horowitz:2010gk}.
Thus, the AdS/CFT correspondence provides a perfect playground to explore the role of disorder within a model
for strongly interacting superconductivity. This is precisely what we do in this manuscript by promoting the
chemical potential in the holographic superconductor to a random space-dependent function. The main rationale for this choice of disorder relies on the fact that the chemical potential defines the local energy of a charged carrier placed at a given position $x$ coupling with the particle number  $n(x)$ locally. Therefore, our choice of disorder replicates a local disorder in the on-site energy.  This is the simplest protocol one would implement. Moreover, once disorder is introduced in such an interacting system, all obervables will become disordered and, therefore, the physics is not expected to depend on the way disorder is implemented. 

In this paper we focus on two aspects.  The first one is the effect of disorder on the critical temperature for superconductivity: we find that the critical temperature for setting the superconductivity is \emph{increased} by the presence of the disorder. The second aspect is the study of certain universality of the power spectra of the condensate  and charge density as functions of the power spectrum of the signal defining the noise. Namely, for a given random signal with power spectrum of the form $k^{-2\alpha}$ we study the power spectrum of the condensate  $k^{-2\Delta(\alpha)}$ and of the charge density $k^{-2\Gamma(\alpha)}$ and report some interesting universal behavior. We interpret this behavior as a particular form of renormalization of small wave-lengths.

{\it Noisy holographic supeconductor.}--
To build a noisy holographic $s$-wave superconductor in 2+1 dimensions we start with the action introduced originally in \cite{Hartnoll:2008vx,Hartnoll:2008kx}. Namely, we consider the dynamics of a Maxwell field and a charged scalar in a fixed metric background:

\be
S=\int d^4 x\,\sqrt{-g}\left(-{1\over4}F_{\mu\nu}\,F^{\mu\nu}-(D_\mu\Psi)(D^\mu\Psi)^\dagger-m^2\Psi^\dagger\Psi
\right)\,.
\ee
The system is studied on the  Schwarzschild-AdS metric:
\bea
ds^2&=&{1\over z^2}\left(-f(z)dt^2+{dz^2\over f(z)}+dx^2+dy^2
\right),\nonumber \\
f(z)&=&1-z^3\,,
\eea
where we have set the radius of AdS, $R=1$, and the horizon at $z_h=1$.
Let us now take the following (consistent) Ansatz for the matter fields:
\bea
&&\Psi(x,z)=\psi(x,z)\,,\quad \psi(x,z)\in {\mathbb R}\,,\\
&&A=\phi(x,z)\,dt\,.
\eea
The resulting equations of motion read:
\bea
 \hspace*{-0.5cm} &&\partial_z^2\phi+\frac{1}{f}\,\partial_x^2\phi-{2\psi^2\over z^2\,f}\,\phi=0\,,\label{eomphi}\\
 \hspace*{-0.5cm} &&\partial_z^2\psi+\frac{1}{f}\,\partial_x^2\psi+\left({f'\over f}-{2\over z}\right)\partial_z\psi+\frac{1}{f^2}\left(\phi^2-{m^2\,f\over z^2}
\right)\psi=0\,. \nonumber\\ \label{eompsi}
\eea
In what follows we will choose the scalar to have $m^2 = -2$, corresponding to a dual operator of dimension 2.

We shall first study the UV asymptotics of equations (\ref{eomphi}), (\ref{eompsi}). Near $z=0$ their solution is given by:
\bea
&&\phi(x,z)=\mu(x)+\rho(x)\,z+\phi^{(2)}(x)\,z^2+o(z^3)\,,\\
&&\psi(x,z)=\psi^{(1)}(x)\,z+\psi^{(2)}(x)\,z^2+o(z^3)\,,
\eea
where $\mu(x)$ and $\rho(x)$ correspond to space-dependent chemical potential and charge density respectively. The functions
$\psi^{(1)}(x)$ and $\psi^{(2)}(x)$ are identified, under the duality, with the source and VEV of an operator of dimension 2. Subsequent $x$-dependent coefficients in this small $z$ asymptotic expansion are determined in terms of $\mu(x),\;\rho(x),\;\psi^{(1)}(x)$ and $\psi^{(2)}(x)$.
In the IR, requiring regularity implies that $A_t$ vanishes at the horizon. Hence, we consider an asymptotic expansion of the form
\bea
 \hspace*{-0.5cm} &&\phi(x,z)=(1-z)\,\phi_h^{(1)}(x)+(1-z)^2\,\phi_h^{(2)}(x)+\ldots \,,\nonumber \label{phiIRexp}  \\
 \hspace*{-0.5cm} &&\psi(x,z)=\psi_h^{(0)}(x)+(1-z)\,\psi_h^{(1)}(x)+(1-z)^2\,\psi_h^{(2)}(x)+\ldots\,,\nonumber \label{psiIRexp}
\eea
where the ellipsis stands for higher order terms.

By redefining the scalar as:
\be
\chi(x,z) = {1-z\over z}\,\psi(x,z)\,,
\ee
%
we arrive at the following boundary value problem:
\bea
\chi(x,0)&=&0, \qquad  \phi(x,0)=\mu(x)\,,\; {\rm UV}\,\,\,  z\to 0\,, \nonumber \\
\chi(x,1)&=&0\;,\qquad \phi(x,1)=0\,,\;\;\;\;\; {\rm IR}\,\,\, z\to 1\,.\label{bcirpsi}
\eea

This choice of boundary conditions corresponds to spontaneous breaking of the $U(1)$ symmetry with order
parameter $\langle {\cal O}\rangle\propto\psi^{(2)}(x)$. From now on we use the angle brackets asociated with  ${\cal O}$ exclusively to refer to average over $x$.

{\it Introducing disorder.}--
We are interested in solving the system given by equations (\ref{eomphi}) and  (\ref{eompsi}) in the
presence of disorder. Let us take the following form for the noisy chemical potential:
\bea
\nonumber
\mu(x)&=&\mu_0+\epsilon\sum_{k=k_0}^{k_*}{\sqrt{S_k}}\,\cos(k\,x+\delta_k)=\\
&=&\mu_0+\epsilon\sum_{k=k_0}^{k_*}{1\over k^\alpha}\,\cos(k\,x+\delta_k)\,,
\label{noisefunc}
\eea
where  $S_k$ is the power spectrum and $\delta_k\in [0,2\pi]$ are random phases. Ensemble averages means averaging over these i.i.d.\ phases. Unless stated otherwise, we consider $\alpha=1$. This means that our noise will be continuous but without well defined derivatives in the limit $k^*\to\infty$. The correlation function of the noise is
\begin{equation}
\left<\mu(x')\mu(x)\right>-\mu_0^2=\sum_{k=k_0}^{k*}\frac{1}{k^2}\cos(k(x'-x)).
\end{equation}
We observe then that, in the limit $k^*\to\infty$, $k_0$ defines the inverse of the correlation length.

We discretize the space, and impose periodic boundary conditions in the $x$ direction, leading to $k$ with values:
\be
k_n={2\pi\,n\over L}\quad {\rm with}\quad
1\leq n\leq{L\over a}-1\,,
\ee
where $L$ is the length in the $x$ direction of our cylindrical space, and $a$ is the lattice spacing in $x$. Note that there is an IR scale given by $k_0$ and a UV scale defined by $k_*=\frac{2\pi}{a}$.  In the limit of a large number of modes, the form of Eq. (\ref{noisefunc}) tends to a Gaussian distributed random function; also note that the power of $1/k$ determines the differentiability properties of $\mu(x)$. We also considered other realizations of disorder, for example,
\be
\mu(x)=\mu_0+ \epsilon(x),
\label{noisefunc2}
\ee
where $\epsilon(x)$ is a random function in $(-W,W)$ taking i.i.d.\ values at different lattice sites (with $W\le \mu_0$). This is the extreme limit in which correlation is lost at a lattice site distance.

We have solved the original system of partial differential equations (\ref{eomphi}) and  (\ref{eompsi}) with boundary conditions (\ref{bcirpsi}) using finite difference with a second order central scheme. Most of the simulations were done independently in Mathematica and in Python. The latter ones ran in the University of Michigan Flux cluster. Our typical result contains a grid of $ 100\times 100$ points but we have gone up to $200\times 200$ to control issues of convergence and optimization. We used a relaxation algorithm to search for the solution and use an ${\cal L}_2$ measure for convergence which in most cases reached $10^{-16}$. As the source of randomness we used $\mu(x)$ given by (\ref{noisefunc}) and also (\ref{noisefunc2}) for uniform and Gaussian distributions.

The scales involved in the problem are: AdS radius $R$, black hole temperature (horizon position) $z_h$ (set to 1), 
chemical potential $\mu_0$ and noise scale $k_0$. There is a distinction between the noise scale and the strength 
of the noise (variance of $\mu(x)$) which we parametrized by $\epsilon$. In the realization (\ref{noisefunc}) we 
have $k_0$ and $\epsilon$. In the context of the realization (\ref{noisefunc2}) when $\epsilon(x)$ takes values in 
a uniform distribution in a certain interval,  the discretization of the problem leads to the introduction of a 
cell in the grid, say of length $a$ along $x$. In this case assigning random values to $\mu(x)$ essentially amounts 
to introducing a correlation length, equal to $a$, such that inside this length the chemical potential is 
correlated but outside this length it is not. Note, however that the IR scale in this choice continues to be $L$ 
which is related to $k_0$ above and the UV scale is $a$ which is related to $k_*$ above. For the realization of the 
noise following Eq. (\ref{noisefunc}) we characterized the strength of disorder by introducing $w$ as 
$\epsilon=\frac{2}{5}\mu_0(w/10)$,  where $w=0$ corresponds to the homogeneous case and $w=10$ is the largest 
$\epsilon_{max}=\frac{2}{5} \mu_0$.We choose this maximum value of the strength by demanding that $\mu(x)$  
remains positive. This means that our analysis is restricted to a relatively small window of disorder. It might happen that
as for some models of strongly coupled superfluids \cite{Dang,Chandra} the superconducting phase is distroyed for
large enough disorder.


For the noise realization given by Eq. \ref{noisefunc2}, the maximum value $\epsilon_{max}$ is $\mu_0$ and the dimensionless measure of the strength of disorder is proportional to $W/\mu_0$. Our definition of $w$ corresponds, in the standard solid state notation, to $1/k_F l$, where $k_F$ is the Fermi momentum and $l$ is the mean free path \cite{PP,s-wave}.  With this choice of scales we have essentially two dimensionless quantities $T/\mu_0$ and the strength of the noise $w$. The results we present below are most marked for the realization given by Eq. (\ref{noisefunc}) but realizations in  Eq. (\ref{noisefunc2}) show similar behavior.

\begin{figure}[htp]
\begin{center}
\includegraphics[width=3.7in]{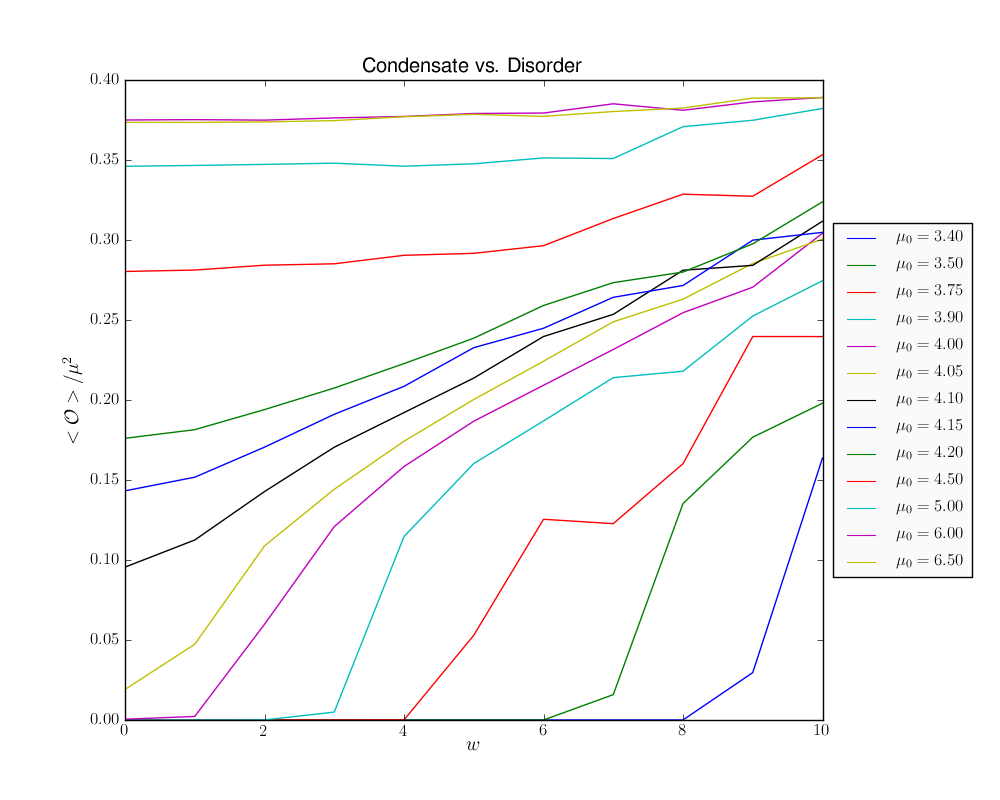}
\caption{\label{CondensatevNoise} Average of the condensate as a function of the strength of disorder using $k_0=1$. The value of the condensate grows with increasing disorder strength, $w$.}
\end{center}
\end{figure}

{\it Results.}-- An intuitive way to summarize our results is presented in Fig. (\ref{CondensatevNoise}) where we
track the normalized value of the condensate $\langle {\cal O}\rangle/ \mu^2$ as a function of the dimensionless
strength of disorder, $w$. Here $\langle \dots \rangle$ stands for average in $x$.
A key feature of this diagram is that increasing disorder leads to a nonzero condensate in situations where the
homogeneous case does not condense. The plot in Fig.\ \ref{CondensatevNoise} was made considering one realization
of the noise. We have performed a preliminary analysis of the variance produced by the intrinsic randomness of the
functions, namely, we have in some cases considered many realizations of the same noise $(\mu_0,k_0,k_*)$  in 
equation (\ref{noisefunc}). We have found that up to $\mu_0$ around $4.00$ this effect is negligible, however for smaller values of $\mu_0$ and with increasing values of the noise strength, $w$, the error becomes significant.  This enhancement of superconductivity is one of our main results.

\begin{figure}[htp]
\begin{center}
\includegraphics[width=3.7in]{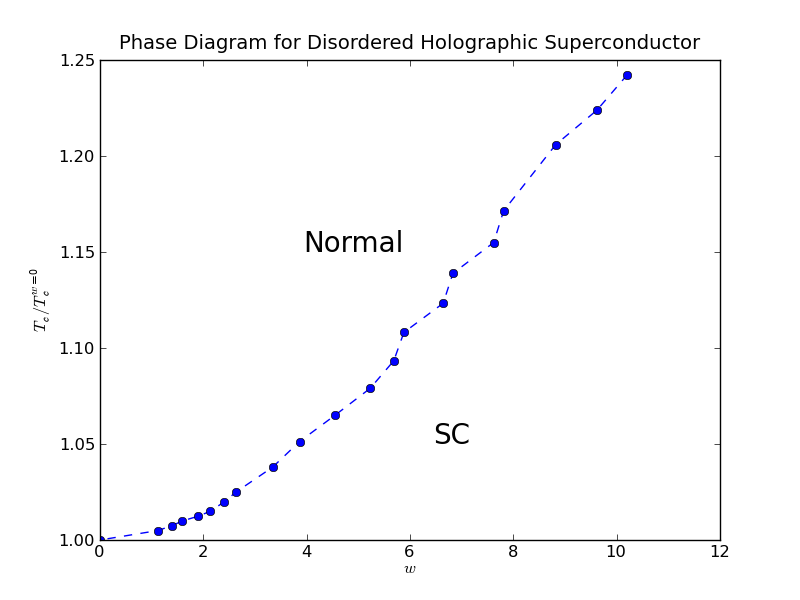}
\caption{\label{CriticalTemperature} Phase diagram, dependence of the critical temperature on the strength of the noise.}
\end{center}
\end{figure}

In a separate plot we explore, in more detail, the change of critical temperature $T_c$ with the strength of
the noise $w$. Namely, we set a value ($25\%$ of the maximum) for which we consider the condensate to be clearly
nonzero and find the smallest value of the homogeneous chemical potential for which this value of disorder $w$ leads
to a nonzero condensate. The results are plotted in figure \ref{CriticalTemperature} which represent our version of the phase diagram of disorderd superconductors as presented, for example, in \cite{s-wave} for realistic $s$-wave superconductors. We will briefly compare our results with the literature in the conclusions.

We note that the parameter $k_0$ affects the results numerically. Recall that the scale $k_0$ is related to the compactification length of the $x$-coordinate. What ultimately matters is the value of this length $L$ with respect to the range in the $z$-coordinate which is fixed to one. The simulations used to generate Figs. \ref{CondensatevNoise} and \ref{CriticalTemperature} used $k_0=1$. For example, taking $k_0=2\pi$, leads to a more modest increment in the value of the condensate. Similarly, considering noise realizations of the form given in Eq. (\ref{noisefunc2}) leads to qualitatively similar results but the numerical enhancement is more modest.

\begin{figure}[htp]
\begin{center}
\includegraphics[width=1.68in]{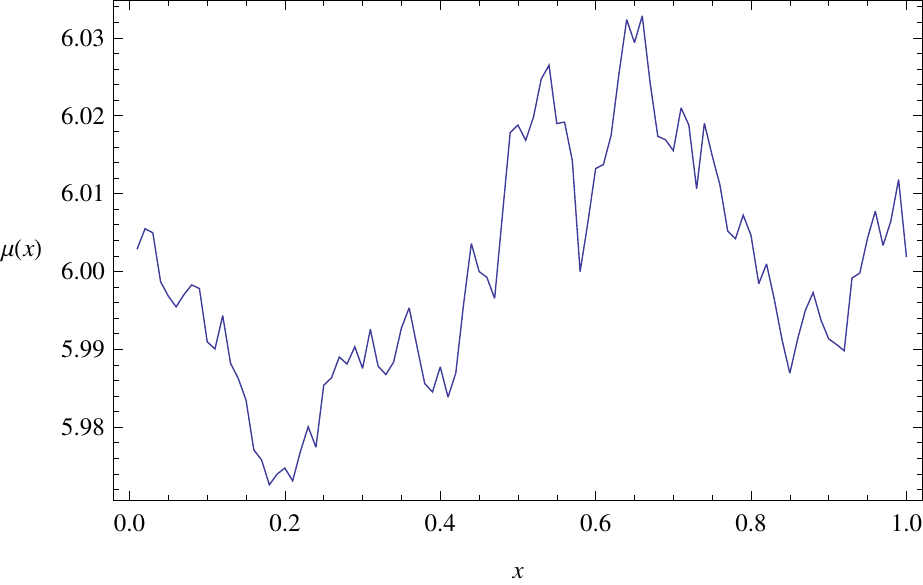}
\includegraphics[width=1.68in]{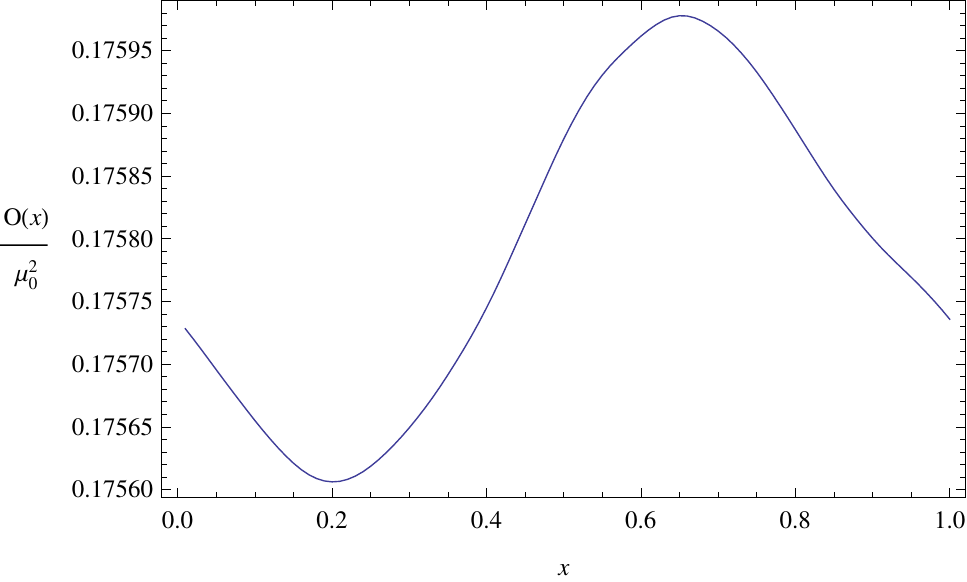}
\caption{\label{Smoothing} Initial chemical potential profile $\mu(x)=6.0+0.1 \sum\limits_{n=1}^{100} \frac{1}{2\pi n}\cos (2\pi n x +\delta_n)$ (left panel) and the corresponding condensate profile (right panel).}
\end{center}
\end{figure}

\begin{figure}[htp]
\begin{center}
\includegraphics[width=1.68in]{muvxfr.pdf}
\includegraphics[width=1.68in]{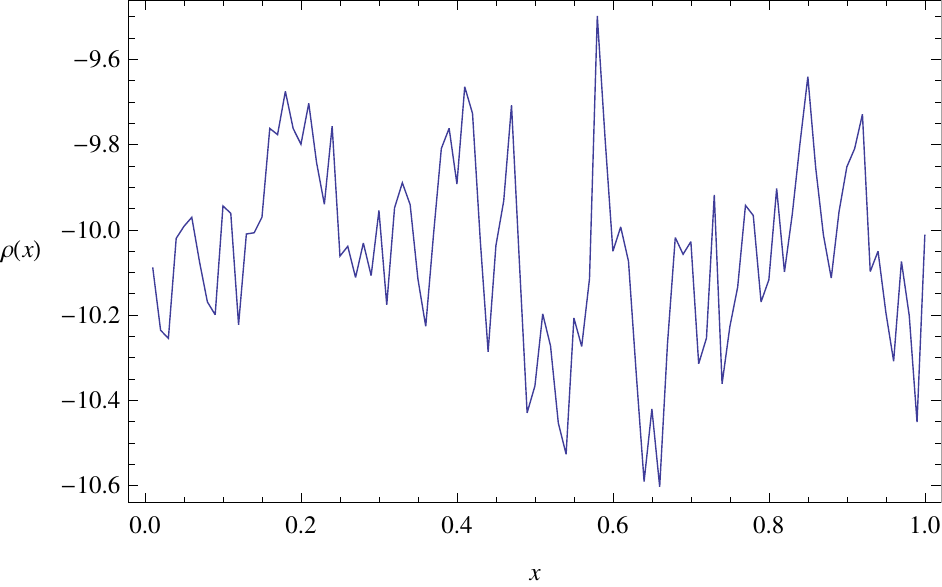}
\caption{\label{Roughening} Initial chemical potential profile $\mu(x)=6.0+0.1 \sum\limits_{n=1}^{100} \frac{1}{2\pi n}\cos (2\pi n x +\delta_n)$ (left panel) and the corresponding charge density profile (right panel).}
\end{center}
\end{figure}

One result that is rather generic is that for a highly irregular chemical potential, we find a very smooth dependence
of the condensate on the coordinate $x$. 
A typical form of $\mu(x)$ and its corresponding ${\cal O}(x)$ are represented in Fig. (\ref{Smoothing}).
On the other hand, the opposite happens for the charge density. A noisy chemical potential will translate into an
even noisier charge density. A typical form of $\mu(x)$ and its corresponding $\rho(x)$ are represented in
Fig. (\ref{Roughening}).

This kind of smoothing/roughening points to a renormalization of sorts, where higher harmonics in ${\cal O}$ are 
suppressed with respect to their spectral weight in $\mu$. Here we pursue this idea further. To characterize this 
renormalization quantitatively we consider a boundary chemical potential of the form presented in equation 
(\ref{noisefunc}) but considering now different values for $\alpha$.
The choice of this parameter $\alpha$ determines the degree of differentiability (smoothness) of the initial 
profile. To make the concept of renormalization more precise we consider the power spectrum of the signal $\mu(x)$ 
which is essentially proportional to $k^{-2\alpha}$; we also consider the power spectrum of the condensate 
${\cal O}(x)$ which we find numerically well approximated by $k^{-2\Delta}$. We find that 
$\Delta\simeq 1.9+1.0\alpha$ is therefore larger than $\alpha$ signaling that the weight of the high-$k$ harmonics 
is smaller in ${\cal O}$ than in $\mu$. Similarly, we approximate the power spectrum of the charge density 
$\rho(x)$ as $k^{-2\Gamma(\alpha)}$, but this time we find $\Gamma<\alpha$. Below we present a plot of $\Delta$ 
and $\Gamma$ versus $\alpha$ for a wide range of values. The error bars were computed based on considering many 
realizations of the given noise.

\begin{figure}[htp]
\begin{center}
\includegraphics[width=1.68in]{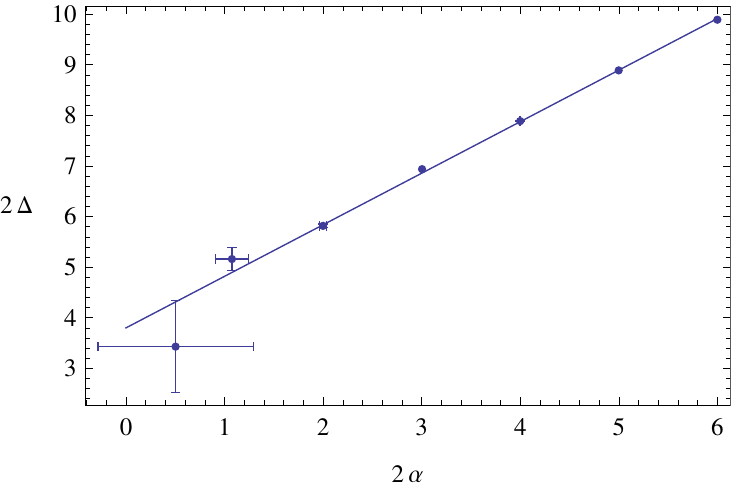}
\includegraphics[width=1.68in]{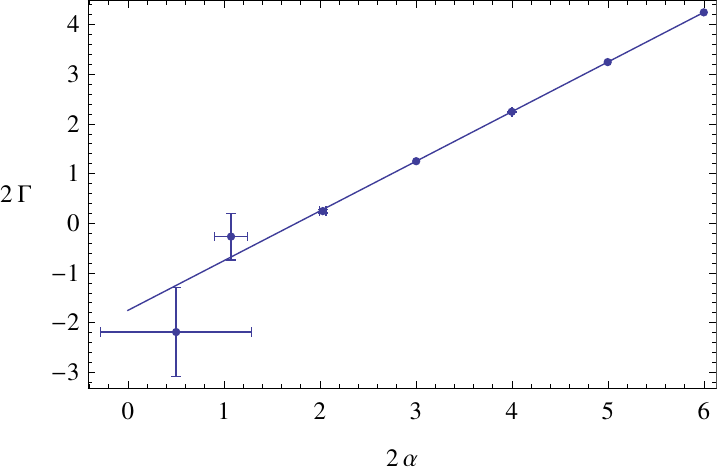}
\caption{\label{RGToy} Renormalization of the disorder: Condensate $\Delta=1.9+1.0\,\alpha$ (left panel) and charge
density $\Gamma= -0.88+ 1.0\,\alpha $ (right panel). This plot was made considering $k_0=2 \pi$, $\mu_0=6$ and
$\epsilon=1$. We can see that the flater the spectrum, the larger the error. This might be associated with the fact
that large momenta are more sensitive to the discretization. }
\end{center}
\end{figure}

We have gathered evidence that this behavior is also rather independent of the value of the mass of the scalar field, or, in the field theory language, it is independent of the conformal dimension of the order parameter. For example, we have confirmed a similar behavior for $m^2=0$. This universality of RG is one of the main observations of our work and its origin seems to be in the strongly coupled nature of the problem. The weak field theory intuition would dictate that $\Delta$ should be well approximated by the conformal dimension associated with the order parameter and here we verify that it is not.

It is also interesting to point out that this behavior does not depend on any of the parameters of our theory,
i.e. $k_0$, $\mu_0$ or $\epsilon$. This means that we can redo Fig. (\ref{RGToy}) for the charge density in the
normal phase. This particular case is interesting, since the theory becomes linear and we can therefore separate
variables. Being that the case, we can recompute the power spectrum solving the equations of motion using a simple
$Mathematica$'s NDSolve command  and we get $\Gamma= -1.0+ 1.0\,\alpha$.
The slope of this fit agrees with that found both for the condensate and the charge density in the broken phase
(Fig. (\ref{RGToy}))
\footnote{A similar behavior was observed in \cite{Hartnoll:2014cua} for the spectrum of a scalar operator.}.
As for the ordinate we expect the aparent discrepancy to vanish for numerics with thinner grids.


{\it Conclusions.}-- In this paper we report two interesting findings: (i) The critical temperature of holographic superconductors increases with the increase in the strength of disorder and (ii) The power spectrum of the condensate and charge density are governed by fairly universal relations depending on the power spectrum of the original random signal.

Let us cautiously compare our results with the situation in the condensed matter literature. In the condensed 
matter literature about superconductors some results point to a degrading of $T_c$ with the strength of the disorder
\cite{Kapitulnik}, 
\cite{s-wave}. Other results, however, point to an enhancement of $T_c$ \cite{Fisher-Fisher-Huse}. The precise role 
of the interactions in these studies is hard to gauge. 
Our results clearly point to an enhancement of $T_c$ but we 
should warn the reader that the role of interactions in our context is central and direct comparison with previous 
studies in the condensed matter literature will require a considerable amount of work likely at the level of 
\cite{BAA} for dirty superconductors, that is, an analysis able to sum the electron-electron perturbation theory to 
all orders.

It might be more pertinent to compare our results with the literature for strongly coupled disordered superfluids.
Indeed the numerical simulations of \cite{Dang, Chandra} have shown
that disorder can triger an insulator to superfluid phase transition in systems that can be of relevance both for
superfluids and high-Tc superconductors.


It is worth pointing out that other holographic discussions, which could be considered as technical precursors to our work in that they solved simplified versions of our system, seem to also point to an enhancement of $T_c$. For example,  \cite{Bao:2011pa} considered a time-dependent stimulation that lead, for a range of frequencies to increasing $T_c$, see however \cite{Natsuume:2013lfa} (also \cite{Li:2013fhw}).  Spatially modulated chemical potentials considered, for example, in \cite{pajer}, \cite{Erdmenger:2013zaa} (see also \cite{Domokos:2013kha}) point instead to a reduction of $T_c$.

The universality of the result $k^{-\alpha} \mapsto (k^{-\Delta(\alpha)}, k^{-\Gamma(\alpha)})$  for the order parameter and the charge density seems to be a general property of the gravity equations of motion. We have also provided supporting evidence that this result is largely independent of the value of the mass of the scalar field. An immediate and intriguing conclusion is that the operator mostly responsible for the RG is not necessarily the one to which the scalar field couples. This ``universality'' is interesting in the framework of the AdS/CFT and deserves further investigation.

The enhancement of  $T_c$ with the strength of the noise in holographic superconductors deserves more scrutiny.  It could be an important prediction of holographic superconductivity.
It would be particularly interesting to consider other types of holographic superconductors (like, for instance, $p$-wave \cite{Gubser:2008wv}), and carry out a similar analysis there. We leave this for future investigation.

In this manuscript we focused largely on the behavior of the condensate and the charge density averaged over the $x$ direction. There seems to be a rich structure in the $x$-dependence of such quantities (see Figs. (\ref{Smoothing}) and (\ref{Roughening})). In particular, the condensate seems to show potential islands of superconductivity. It would be interesting to pursue the appearance of islands of superconductivity and the effect of different noises on the conductivity (see \cite{Hartnoll:2012rj}). We hope to address such questions in a separate publication.

{\bf Acknowledgments} We are thankful to C. Herzog, J. Maldacena, D. Musso, J. Sonner, D. Tielas, D. Vaman and C. Wu.  D.A.
and L.PZ.  acknowledge hospitality of the  MCTP  and ICTP, respectively. L.PZ. and A.S. acknowledge the hospitality
of GGI during the gestation of the project. D.A. thanks the FRoGS for unconditional support. This work was supported
by the DoE Grant DE-SC0007859.

\end{document}